# Seismocardiographic Signal Timing with Myocardial Strain

Amirtaha Taebi[1], Richard H. Sandler[1,2], Bahram Kakavand[2], Hansen A. Mansy[1]
[1] Biomedical Acoustics Research Lab, University of Central Florida, Orlando, FL 32816, USA
[2] Nemours Children's Hospital, Orlando, FL 32827, USA
{taebi@knights., hansen.mansy@ }ucf.edu

*Abstract*—Speckle Tracking Echocardiography (STE) is a relatively new method for cardiac function evaluation. In the current study, STE was used to investigate the timing of heart-induced mostly subaudible (i.e., below the frequency limit of human hearing) chest-wall vibrations in relation to the longitudinal myocardial strain. Such an approach may help elucidate the genesis of these vibrations, thereby improving their diagnostic value.

*Introduction*—Seismocardiograms (SCG) are the cardiac vibrations detected at the chest wall surface [1]–[3]. The relation between SCG waves and cardiac activity are not fully understood [4]. However, it is believed that mechanical processes such as myocardial contraction, blood momentum changes, and valve closure are sources of these vibrations [5]–[8]. STE was used in this study to help explain the relationship between myocardial contraction strain and SCG chest wall vibration signal morphology.

*Methods*—After informed consent, the SCG signal from a healthy 18 year old male was captured using a uniaxial accelerometer (352C65, PCB Piezotronics, Depew, NY). The SCG sensor was placed at the left lower sternal border at the level of the 4th intercostal space. Electrocardiographic (ECG) and respiration flow rate signals were recorded simultaneously using a control module (IX-TA-220, iWorx Systems, Inc., Dover, NH). The STE was recorded at the same time (EPIQ 5, Philips, Netherlands). The ECG tracing was simultaneously acquired by the echocardiography machine as well for signal synchronization. The SCG events were grouped into inspiratory and expiratory events (as measured by positive and negative flow rates, respectively). These events were then aligned in time with the left ventricular (LV) strain curve supplied by the STE technique.

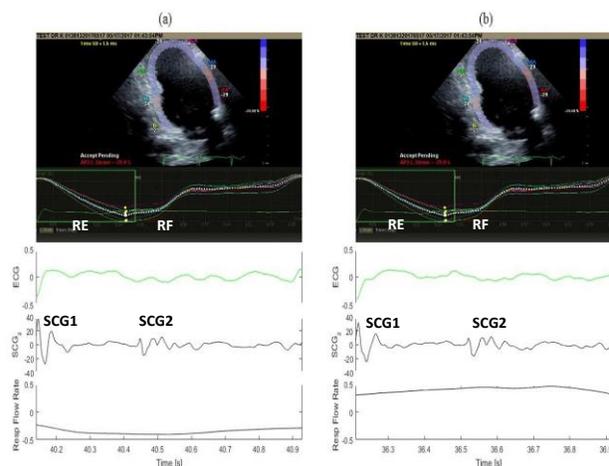

Fig. 1. Simultaneously acquired STE, ECG and SCG signals during (a) expiration, and (b) inspiration, from one subject. SCG1 and SCG2 starts before rapid ejection (RE) and Rapid filling (RF), respectively.

*Results*—Fig. 1 shows the myocardial strain curves for a representative expiratory and inspiratory SCG cycle. Small changes in the SCG waveform were seen with respiration. The strain in Fig. 1 was negative and is defined as the fractional change in the heart tissue dimension in comparison with the original tissue dimension. The SCG waveform contained two major events, SCG1 and SCG2, which occurred close to the first and second heart sound timing, respectively. The changes in longitudinal strain was due to the well-known LV twisting motion. The SCG signal peak (at SCG1 maximum) occurred shortly after the ECG R wave and during the early part of rapid ejection (RE). During ejection, the apex rotated counterclockwise while the base performed a clockwise rotation. This resulted in a decrease in the longitudinal length of the LV. The myocardial strain increased (i.e. became more negative) and reached a plateau (isovolumic relaxation) before the beginning of SCG2. Towards the end of SCG2, the strain started to be less negative and rapid filling (RF) took place.

*Conclusions*—The SCG wave timings were compared with the myocardial strain curve for the first time. The results showed that the heart muscle experienced lowest negative mechanical strain around SCG1, with

Research reported in this publication was supported by the National Institutes of Health under R44HL099053.

the highest negative strain occurring just prior to SCG2. STE might help characterize the LV (un-)twist patterns on the SCG waveforms which might lead to distinguishing normal SCG signals from abnormal ones. More studies are needed to investigate the differences in SCG morphology based on heart muscle contractile state in both healthy subjects and those with cardiovascular disease.